\documentclass[12pt,a4paper]{article}
\usepackage[german, english]{babel}
\usepackage{a4wide}
\usepackage{amsmath}
\usepackage{amssymb}
\usepackage{ifthen}
\usepackage{epsfig}
\newcounter{fig}   \newcommand{\lbfig}[1]{\refstepcounter{fig}
\label{#1} }

   \newcommand{\diag}{\operatorname{diag}}

\newcommand{\Gr}[2][]{\ensuremath{ \mathrm{#2}
\ifthenelse{\equal{#1}{}} {} {(#1)} } } \newcommand{\lie}{\mathfrak}
\newcommand{\gr}[2][]{\ensuremath{ \lie{#2} \ifthenelse{\equal{#1}{}}
{} {(#1)} } }     
\newcommand{\minuspt}{\\[-4pt]}
\newcommand{\abb}[3]{\ensuremath{ \ifthenelse{\equal{#1}{}}{}{#1:} #2
\mapsto #3}} 
\newcommand{\abbildung}[6][]{\begin{align}
\ifthenelse{\equal{#1}{}}{}{\label{#1}}
\ifthenelse{\equal{#2}{}}{}{#2:} #3 & \rightarrow   #4  \minuspt #5 &
\mapsto #6  \notag \end{align}} 
\newcommand{\bea}{\begin{eqnarray}}
\newcommand{\eea}{\end{eqnarray}}

\def\gl22{\mathop{\mathfrak{gl}(2|2)}\nolimits}

\def\diag{{\rm diag}}

\def\Tr{{\rm Tr}}

\def\sq3{\sqrt{3}}
\def\isq3{\frac{1}{\sqrt{3}}}

\def\t2{{\Bbb T}^2}
\newcommand\grsim{\mathrel{\hbox{\lower1ex\hbox{\rlap{$\sim$}\raise1ex\hbox{$>$}}}}}
\newcommand\losim{\mathrel{\hbox{\lower1ex\hbox{\rlap{$\sim$}\raise1ex\hbox{$<$}}}}}
\newcommand {\be} {\begin{equation}}
\newcommand {\ee} {\end{equation}}

\newcommand{\half}{\frac{1}{2}}

\begin{document}

\begin{titlepage}
\vspace*{2cm}

\begin{center}
{\bf\large Properties of non-BPS $SU(3)$ monopoles}
\vspace{2.0cm}

{\sc\large Ya. Shnir}\\[12pt]
{\it  Institute of Physics, University of Oldenburg}\\
{\it D-26111, Oldenburg, Germany}

\end{center}

\date{~}

\bigskip

\begin{abstract}
This paper is concerned with magnetic monopole solutions 
of $SU(3)$ Yang-Mills-Higgs system beyond the Bogo\-mol'nyi-Pra\-sad-Sommer\-field 
limit. The different $SU(2)$ embeddings, which correspond to the fundamental monopoles,
as well the embedding along composite root are studied. The interaction of two different 
fundamental monopoles is considered. Dissolution of a single fundamental 
non-BPS $SU(3)$ monopole in the limit of the minimal symmetry breaking is analysed.    
\end{abstract}
\bigskip


\end{titlepage}
\section{Introduction}
One of the most interesting direction of the modern field theory is related with 
study of magnetic monopoles in spontaneously broken gauge theories.
It was shown almost immediately after  
discovering of the celebrated 't~Hooft-Polyakov monopole solution  
of the simple $SU(2)$ Yang-Mills-Higgs theory \cite{Hooft74,Polyakov74},  
that existence of such monopole field configurations    
is a generic prediction of grand unified theories.

More complicated is the question of existence of multimonopole solutions 
with topological charge $n>1$. Here most progress have been made in 
the original   $SU(2)$  theory, where, in   
the limit of vanishing scalar potential (so called 
Bogo\-mol\-nyi-Pra\-sad-Som\-mer\-fi\-eld (BPS) limit), the monopoles
satisfy the first order equations. The latter are just 3-dimensional reduction 
of the integrable self-duality equations. However there is no analytical solution 
of the Yang-Mills-Higgs model which would correspond to a system of two 
separated  't~Hooft-Polyakov monopoles in a general case of non-vanishing Higgs 
potential.  

There are arguments that 
such a system corresponds to a saddle point of the energy functional of the
Yang--Mills--Higgs system \cite{Ward81,Taubes82}, not a minimum as in the
single-monopole case.  Therefore, this configuration in not static and 
we have to take into account the effect of the interaction between the 
monopoles which was analysed in \cite{Goldberg,Kunz98,KisSh97}.

It would be rather difficult to find an
exact description of the time evolution of such a system in the general case.
Considering the $SU(2)$ Yang-Mill-Higgs system, Taubes proved that 
the magnetic dipole solution could exists \cite{Taubes82}.  
The detailed consideration of this axially symmetric  
$SU(2)$ monopole-antimonopole pair configuration 
was given in \cite{Rueber85,Kunz00}. Recently, more general static equilibrium solutions 
have been constructed, representing chains, where $m$ monopoles and antimonopoles alternate
along the symmetry axis \cite{KKS}.

Clearly, first step beyond the simplest $SU(2)$ gauge theory 
is to consider an extended $SU(3)$ model. The properties of the 
corresponding monopoles, which generalise the spherically symmetric 
't Hooft--Polyakov solution, were discussed in \cite{Chackra75,Marciano75}.
This solution is actually a simple embedding of the  $SU(2)$ 
monopole into corresponding Cartan subalgebra of the $SU(3)$ model. 
However the pattern of symmetry breaking of latter theory is 
different. One have to separate two different situations: the minimal symmetry
breaking $SU(3) \to U(2)$ and the maximal symmetry breaking:  
$SU(3) \to U(1)\times U(1)$. Moreover, the 2-monopole solution arises in this
picture on a very natural way as a configuration which corresponds to the 
composite root of the Cartan-Weyl basis. 

Detailed analyse of  the BPS 
monopole solutions in the gauge theory with a large gauge group by  
E.~Weinberg \cite{Weinberg79,Weinberg80,Weinberg82} 
unexpectedly shows that some of these solutions correspond
to massless monopoles, so called cloud. The moduli space approximation allows to study 
the interaction energy of well separated BPS monopoles and the formation of the 
non-Abelian cloud \cite{Irwin97,Lu98}.   
However there is very little information about properties of non-BPS $SU(3)$ 
multi-monopoles, their interaction and behaviour in the massless limit since 
the powerful Nahm formalism cannot be applied in that case.   

If we restrict our discussion to the system of two monopoles,  
we could expect that, alongside with a straightforward 
embedding of $SU(2)$ spherically symmetric mo\-no\-pole 
configuration  along a given positive simple root of the Cartan-Weyl basis 
\cite{Corrigan76,Kunz87}, there are some other solutions. 
The latter  in the BPS limit would correspond to the distinct 
fundamental monopoles which are embedded along composite roots. 

Clearly, the character of the interaction between these
monopoles depends from their relative orientation in the group space and it 
has to be quite different from the interaction of the monopoles, 
which are embedded along the same simple root of the 
Cartan-Weyl basis. Furthermore, if the symmetry is breaked minimally, there 
should be a counterpart of the massless BPS monopole whose existence was recently 
proved from the Hahm formalism \cite{Irwin97,Lu98}. 

In the present note we discuss the spectrum of the spherically symmetric 
non-BPS monopole solutions of the $SU(3)$ Yang-Mill-Higgs theory and analyse 
the properties of the solution with minimal symmetry breaking.   
     
\section{Cartan-Weyl basis of  $SU(3)$ group}
We consider the $SU(3)$ Yang-Mills-Higgs Lagrangian density
\bea           \label{LagrGG-8}
-L &=& \frac{1}{2} \Tr\, F_{\mu\nu}F^{\mu\nu} + \Tr\, D_\mu \Phi D^\mu \Phi +
V(\Phi)\\ 
&=&
\frac{1}{4} F_{\mu \nu}^a F^{a \mu \nu} + \frac{1}{2} (D^{\mu} \Phi ^a)
(D_{\mu} \Phi ^a )
 + V(\Phi),\nonumber  
\eea
where $F_{\mu\nu} = F_{\mu \nu}^a T^a; ~\Phi = \Phi^a T^a$ and the symmetry 
breaking Higgs potential is\footnote{Most general $SU(3)$-symmetry breaking 
scalar potential can be written in a different form \cite{Burzlaff81,Kunz87}. 
To set a correspondence with the related discussion of the BPS monopoles we shall take 
into account the pattern of the symmetry breaking on a different way.} 
\begin{equation}        \label{pot-8}
V(\Phi) = {\lambda} (|\Phi|^2 - 1)^2.
\end{equation} 
where we used the $su(3)$ norm $|\Phi|^2 = \Phi^a \Phi^a$. 

The $su(3)$ Lie algebra is given by a set of traceless matrices $T^a = \lambda^a/2$ 
where $\lambda^a$ are the standard Gell-Mann matrices, that is we use the normalisation 
$\Tr\, T^a T^b = \half \delta^{ab}$. The structure constants 
are $f^{abc} = \frac{1}{4} \Tr [\lambda^a, \lambda^b]
\lambda^c$ and in the adjoint representation $(T^a)_{bc} = f^{abc}$.

Discussion of monopoles in a gauge theory of higher rank is closely connected with notion of the 
Cartan-Weyl basis \cite{Weinberg79,Weinberg80,Weinberg82}. Let us briefly review the basic elements
of this approach.
 
The 
diagonal, or Cartan subalgebra of $SU(3)$ is given by two generators
\begin{equation}    \label{H-SU3-ch8}
H_1 \equiv T^3 = \frac{1}{2} \left(\begin{array}{ccc}
1&0&0\\
0&-1&0\\
0&0&0\\
\end{array}\right);\qquad 
H_2 \equiv T^8 = \frac{1}{2 \sqrt 3}\left(\begin{array}{ccc}
1&0&0\\
0&1&0\\
0&0&-2\\
\end{array}\right)
\end{equation}
which are composed into the vector ${\vec H} = (H_1, H_2)$. 
The Cartan-Weyl basis of the  $SU(3)$ group can be constructed by addition 
to the commuting elements ${\vec H}$ two raising and lowering generators 
 $E_{\pm{\vec \beta}}$, each for one of two simple roots ${\vec \beta} = 
({\vec \beta}_1,{\vec \beta}_2)$:
 \begin{equation}
[H_i,E_{\vec\beta}] = \beta_i E_{\vec\beta};\qquad  [E_{\vec\beta}, 
E_{-{\vec\beta}}] = 2 {\vec \beta} \cdot {\vec H}
\end{equation} 
%The advantage of this approach is to put a simple Euclidean geometry into
%correspondence to the algebra of Lie group generators.  The roots $\vec
%\beta_i$ correspond to the structure constants of a Lie group. Since the  
%dimension of the $SU(3)$ group is $d=8$, number of positive roots is 3. 
We take the basis of simple roots as (cf. figure \ref{f-03})
\begin{equation}           \label{simple-root}
{\vec \beta}_1 = (1, 0) ;\qquad {\vec \beta}_2 = (-1/2, {\sqrt 3}/2)
\end{equation}
Third positive root is given by the composition of the
first two roots $\vec \beta_3 = \vec \beta_1 + \vec \beta_2 = (1/2, {\sqrt 3}/2)$. 
Note that 
all these roots have a unit length, that is our choice corresponds to the 
self-dual basis: ${\vec \beta}_i^* = {\vec \beta}_i$. 

For any given root ${\vec\beta}_i$ the generators ${\vec \beta}
\cdot {\vec H},~ E_{\pm\beta_i}$ form an $su(2)$ algebra. Let us write 
these generators explicitly in the above defined basis of the simple roots.  
For $\vec \beta_1$ we have 
\bea                   \label{su3-1}
T_{(1)}^3 &=& {\vec \beta}_1 {\vec H} = \frac{1}{2} \left(\begin{array}{ccc}
1&0&0\\
0&-1&0\\
0&0&0\\
\end{array}\right);\\ 
E_{{\vec\beta}_1} &=& \left(\begin{array}{ccc}
0&1&0\\
0&0&0\\
0&0&0\\
\end{array}\right);\qquad 
E_{-{\vec\beta}_1} \equiv \left(\begin{array}{ccc}
0&0&0\\
1&0&0\\
0&0&0\\
\end{array}\right).\nonumber
\eea
For the second simple root ${\vec \beta}_2$ we have
\bea            \label{su3-2}
T_{(2)}^3 &=& {\vec\beta}_2 {\vec H} = \frac{1}{2} \left(\begin{array}{ccc}
0&0&0\\
0&1&0\\
0&0&-1\\
\end{array}\right);\\ 
E_{{\vec\beta}_2}  &=& \left(\begin{array}{ccc}
0&0&0\\
0&0&1\\
0&0&0\\
\end{array}\right);\qquad 
E_{-{\vec\beta}_2}  = \left(\begin{array}{ccc}
0&0&0\\
0&0&0\\
0&1&0\\
\end{array}\right).\nonumber
\eea
The  generators of the $su(2)$ subalgebra which correspond to the  
third  composite root are given by the set of matrices
\bea             \label{su3-3}
T_{(3)}^3 &=& {\vec\beta}_3 {\vec H} = \frac{1}{2} \left(\begin{array}{ccc}
1&0&0\\
0&0&0\\
0&0&-1\\
\end{array}\right);\\ 
E_{{\vec\beta}_3}  &=& \left(\begin{array}{ccc}
0&0&1\\
0&0&0\\
0&0&0\\
\end{array}\right);\qquad 
E_{-{\vec\beta}_3}  = \left(\begin{array}{ccc}
0&0&0\\
0&0&0\\
1&0&0\\
\end{array}\right).\nonumber
\eea
Clearly, the set of matrices $T^a_{(k)}$,  $k=1,2,3$, which includes $ T_{(k)}^3$ of 
Esq (\ref{su3-1}),(\ref{su3-2}) and (\ref{su3-3}), and  
\begin{equation}
T^1_{(k)} = \frac{1}{2}\left(E_{{\vec\beta}_k} + E_{-{\vec\beta}_k}\right), \qquad T^2_{(k)} = 
\frac{1}{2i}\left( E_{{\vec\beta}_k} -  E_{-{\vec\beta}_k}\right)
\end{equation}
satisfy the commutation relations of the $su(2)$ algebras associated with the
simple roots ${\vec \beta}_1, ~{\vec\beta}_2$ and  ${\vec \beta}_3$ respectively.

In other words, the basis of the simple roots $\beta_1$, $\beta_2$ corresponds
to two different ways to embed the $SU(2)$ subgroup into $SU(3)$. Upper
left and lower right $2 \times 2$ blocks are corresponding to the subgroups
generated by the simple roots $\beta_1$ and $\beta_2$ respectively. The
third composite root ${\vec \beta}_3$ generates the $SU(2)$ 
subgroup which lies in the corner elements of the $3\times 3$ matrices 
of the $SU(3)$.  Note there is also so-called maximal embedding, which is given by the set 
of matrices  
\begin{equation}
\begin{split}
{\tilde T}_1 = \frac{1}{\sqrt 2}\left(\begin{array}{ccc}
0&1&0\\
1&0&1\\
0&1&0\\
\end{array}\right);&\qquad 
{\tilde T}_2 = \frac{1}{\sqrt 2}\left(\begin{array}{ccc}
0&-i&0\\
i&0&-i\\
0&i&0\\
\end{array}\right);\nonumber\\
{\tilde T}_3 &= \left(\begin{array}{ccc}
1&0&0\\
0&0&0\\
0&0&-1\\
\end{array}\right)
\end{split}
\end{equation}
These matrices up to an unitary transformation are equivalent to the vector representation 
of $SU(2)$. The very detailed analyse of the corresponding solutions is presented in \cite{Kunz87}.
We shall not consider the maximal embedding in this note.     

\section{$SU(3)$ two-monopole configurations}
%Unlike the original $SU(2)$ `t~Hooft-Polyakov monopole solution, 
%the vacuum manifold ${\cal M}$ of the $SU(3)$ Yang-Mills-Higgs theory is a 
%sphere $S^7_{vac}$ in eight-dimensional space. Thus, the topological
%classification of the solutions is related with mapping of spatial asymptotic
%$S^2$ onto coset space ${\cal M} = SU(3)/H$, where $H$ is a residual symmetry
%of the vacuum. In order to classify the solutions we have to 
%define the unbroken subgroup $H$.  
%
%Note that 
The asymptotic value of the scalar field in some 
fixed direction (e.g. positive $z$-axis) can be chosen to
lie in the Cartan subalgebra
\begin{equation}                      
\Phi_0 =  {\vec h} \cdot {\vec H} ,   
\end{equation}
where ${\vec h} = (h_1, h_2)$ is 2-component vector in the space of Cartan subalgebra. 
Clearly, that is a generalisation of the $SU(2)$ boundary condition 
$\Phi_0 =  \sigma_3/2$. 

Orientation of the Higgs field in the $SU(3)$ root space 
corresponds to the pattern of the symmetry breaking.
If the Higgs vector ${\vec h}$ is orthogonal to none from the simple roots
$\vec \beta_i$, the 
symmetry is maximally broken to the
maximal Abelian torus $U(1)\times U(1)$. 
If inner product of ${\vec h}$ and either of
the simple roots is vanishing, there are two choices of the basis
of simple roots with positive inner product with  ${\vec h}$ which are related
by Weyl reflections. If, for example ${\vec h}$ is orthogonal to $\vec \beta_2$,    
we can choose between two possibilities: $(\vec \beta_1, \vec \beta_2)$ and 
$(\vec \beta_1 + \vec \beta_2, -\vec \beta_2)$.
This is the case of the minimal symmetry breaking $SU(3)\to U(2)$.

Furthermore, the magnetic charge of the $SU(3)$ monopole is also defined as a vector 
in the root space, that is 
$$
g = {\vec g} \cdot {\vec H}
$$
where the vector magnetic charge lies on the dual root lattice\footnote{Recall
that in our self-dual basis $ \vec\beta_i^*  = \vec\beta_i$.}
\cite{GoddOlive77}
\begin{equation}             \label{wein-quant}
{\vec g} =\sum\limits_{i=1}^{r}n_i \vec\beta_i^*  =  {\vec g}_1 +  {\vec g}_2 =
n_1 {\vec \beta_1} + n_2 {\vec \beta_2} 
\end{equation}
where $n_1$ and $n_2$ are integer and ${\vec g}_1, {\vec g}_2$ are the magnetic 
charges associated with the corresponding simple roots. Thus, 
\begin{equation}        \label{su3quant}
g = {\vec g} \cdot {\vec H} = 
\left(n_1 - \frac{n_2}{2}\right) H_1
+ \frac{\sqrt 3}{2} n_2 H_2
\end{equation}

An $SU(3)$ spherically symmetrical monopole configuration  
can be constructed by a simple embedding \cite{Corrigan76,Bais78,Kunz87}. 
The recipe is obvious: we have to choose one of the simple
roots having a positive inner product with the scalar field, for example,
$\vec \beta_1$, and embed the `t Hooft-Polyakov solution into the 
corresponding $SU(2)$ subgroup. For example, embedding into left upper corner  
$SU(2)$ subgroup  
defines the $\beta_1$-monopole which is characterised 
by the vector charge $\vec g =
(1,0)$ while the embedding into lower right corner  
$SU(2)$ subgroup  
defines the $\beta_2$-monopole with  the vector charge $\vec g =
(0,1)$. Similarly, one can embed  
the $SU(2)$ axially symmetric monopole-antimonopole saddle point
configuration  of \cite{Rueber85,Kunz00} which yields the state 
$\vec g =(0,0)$ or two $SU(2)$ monopoles configuration of \cite{Ward81}
which, depending from the root we choose, 
yields the states $\vec g =(2,0)$ or $\vec g =(0,2)$ respectively. 
 
Embedding of the spherically symmetric $SU(2)$ monopole 
along composite root $\beta_3$ gives a $(1,1)$ monopole with the magnetic 
charge 
$$
g = \vec g \cdot  \vec H = \frac{1}{2} H_1 + \frac{\sqrt 3}{2}H_2
$$ 
The analysis based on the index theorem shows
\cite{Weinberg80}, that this configuration is a simple
superposition of two other fundamental solutions and could be continuously
deformed into solution which describes two well separated single $\beta_1$ and $\beta_2$ 
monopoles. 

It is known that the character of interaction between the $SU(3)$ BPS  
monopoles depends from the type of the embedding \cite{Weinberg96-2}.
This is also correct for non-BPS extention. 
Indeed, then there is only long-range electromagnetic field which mediates 
the interaction between
two widely separated non-BPS monopoles, that is they are considered as classical 
point-like particles with magnetic charges $g_i = \vec g_i \cdot \vec H = \vec \beta_i 
\cdot \vec H$. For a non-zero scalar coupling $\lambda$ the contribution of the scalar field 
is exponentially suppressed. 
The energy of the electromagnetic interaction then originates from 
the kinetic term  of the gauge field  
  $\frac{1}{2} \Tr\, F_{\mu\nu} F^{\mu\nu}$ in the Lagrangian
(\ref{LagrGG-8}). Therefore  
an additional factor $\Tr [(\vec \beta_i \cdot \vec H) (\vec \beta_j \cdot
\vec H) ] = (\vec \beta_i \cdot \vec \beta_j) $ appears in the formula for the energy
of electromagnetic interaction. In the case under consideration $(\vec \beta_1 \cdot \vec \beta_2) 
= -\frac{1}{2}$ while $(\vec \beta_i \cdot \vec \beta_i) = 1$. 
This corresponds to 
an attraction of two different fundamental $SU(3)$ monopoles and repulsion of two 
monopoles of the same $SU(2)$ subalgebra due to non-trivial group structure.  
The energy of interaction between the $\beta_1$ and $\beta_2$ monopoles then 
is: $V_{int} = - \frac{ ({\bf r}_1 {\bf r}_2)}{ r_1^3  r_2^3}$. 

We can check this conclusion by making use of an analogy with the classical 
electrodynamics of point-like charges. Let us suppose that both monopoles are located 
on the $z$-axis at the points $(0,0,\pm R)$.

The electromagnetic field of that configuration can be calculated in the Abelian gauge 
where the gauge field become additive \cite{ArFreud}.
If the monopoles are embedded along the same     
simple root, say  $\beta_1$, we can write the components of the gauge field as 
\begin{equation}
A_r = A_\theta =0;\qquad A_\phi = (1 + \cos \theta_1) \frac{\sigma_3^{(1)}}{2} + 
(1 + \cos \theta_2) \frac{\sigma_3^{(1)}}{2}
\end{equation}
Simple calculation yields the components of the electromagnetic field strength tensor 
\begin{equation}  
\begin{split} 
F_{r\theta} = 0;\qquad F_{r \phi} &= rR\sin^2\theta\left(\frac{1}{r_1^3}\, \frac{\sigma_3^{(1)}}{2}
-\frac{1}{r_2^3}\, \frac{\sigma_3^{(1)}}{2}\right);\\
 F_{\theta \phi} &=-r^2\sin\theta\left(\frac{r-R\cos\theta}{r_1^3}\frac{\sigma_3^{(1)}}{2} + 
\frac{r+R\cos\theta}{r_2^3}\frac{\sigma_3^{(1)}}{2}\right), 
\end{split}
\end{equation}
where $r_1$, $r_2$ are the distances of the point $r$ to the points at which monopoles are placed.
The field energy becomes 
\begin{equation}
E = \Tr \left( \frac{1}{r^2 \sin^2\theta}  F_{r \phi}^2 +  \frac{1}{r^4 \sin^2\theta}  
F_{\theta \phi}^2\right) = \frac{1}{2}\left[ \left(\frac{{\bf r}_1}{r_1^3}\right)^2 + 
  \left(\frac{{\bf r}_2}{r_2^3}\right)^2 + \frac{2 ({\bf r}_1 {\bf r}_2)}{ r_1^3  r_2^3}
\right]
\end{equation}
that is the potential energy of the electromagnetic interaction of two $\beta_1$ monopoles 
is repulsive. 
However, for a $\beta_3$ configuration with vector charge $\vec g = (1,1)$ the components of the 
gauge fields are 
\begin{equation}
A_r = A_\theta =0;\qquad A_\phi = (1 + \cos \theta_1) \frac{\sigma_3^{(1)}}{2} + 
(1 + \cos \theta_2) \frac{\sigma_3^{(2)}}{2}
\end{equation}
and, because $\Tr\, \sigma_3^{(1)}\sigma_3^{(2)} = - 1$, the field energy is 
\begin{equation}
E = \frac{1}{2}\left[ \left(\frac{{\bf r}_1}{r_1^3}\right)^2 + 
  \left(\frac{{\bf r}_2}{r_2^3}\right)^2 - \frac{ ({\bf r}_1 {\bf r}_2)}{ r_1^3  r_2^3}
\right]
\end{equation}
that is $\beta_1$ and $\beta_2$ monopoles 
attract each other with a half-force comparing to the case of the repulsion of two $\beta_1$ 
monopoles.  

Note that for a system of two well separated $SU(3)$ dyons the energy of the classic long-range 
electromagnetic interaction also includes the electric part. The electric charges of the $\beta_1$ 
and $\beta_2$ dyons  are defined with respect to different $U(1)$ subgroups, that is 
alongside the magnetic charges 
they are vectors in the root space: $Q_i = q_i(\vec \beta_i \cdot \vec H)$. Thus, the potential 
of interaction of two identical dyons remains proportional to the inner product $(\beta_i \cdot 
\beta_j)$ as in the case of purely magneically charged configuration.

\subsection{Spherically symmetric $SU(3)$ monopole configuration}
In this simplified consideration above we neglected both the structure of the monopole 
core and the contribution of the short-range massive scalar field. 
Such an approximation to the low energy dynamics of the $SU(3)$ monopoles have been applied in 
the moduli space approach \cite{Weinberg96-2}. However the mechanism of the interaction 
becomes more complicated if the symmetry is brocken minimally. Then one of 
the fundamental monopoles is losing its identity as a located field configuration. If this monopole 
would be isolated it would spread out and disappear. But as its core overlaps with the second massive 
monopole, it ceases to expand \cite{Dancer92,Dancer93,Irwin97,Lu98}.   

To analyze the behavior of two distinct fundamental monopole systems in that limit we study 
the spherically symmetric $SU(3)$ monopoles in a more consistent way. In the BPS limit 
our numerical results can  be compared with the consideration of the paper \cite{Lu98} where 
the Nahm formalism was used  to calculate the monopole energy density.  
 
For each simple root $\vec \beta_i$, which defines an $SU(2)$ subgroup 
with corresponding generators $T^a_{(i)}$, we can define an embedded $SU(2)$ monopole 
as \cite{Bais78} 
\begin{equation}  \label{emb-Bais}
A_n = A_n^a T^a_{(i)};\qquad \Phi = \Phi^a T^a_{(i)} + \phi^{(h)},~~{\rm where}~~ \phi^{(h)} = 
\left(\vec h -  {\vec \beta}_i\, 
({\vec h} \cdot {\vec \beta}_i )\right) \vec H
\end{equation}
The additional invariant term $\phi^{(h)}$ is added to the Higgs field to 
satisfy the boundary conditions on the spatial asymptotic. 
In our basis of the simple roots we can write  
\begin{equation}  \label{h-inv}
\begin{split}
\vec \beta_1:~~~& \phi^{(h)} = \frac{h_2}{2\sqrt 3}  \left(\begin{array}{ccc}
1&0&0\\
0&1&0\\
0&0&-2\\
\end{array}\right);\\
\vec \beta_2:~~~& \phi^{(h)} = \frac{1}{4}\left(h_1 + \frac{h_2}{\sqrt 3}\right)  \left(\begin{array}{ccc}
2&0&0\\
0&-1&0\\
0&0&-1\\
\end{array}\right);\\
\vec \beta_3:~~~& \phi^{(h)} = \frac{1}{4}\left(h_1 - \frac{h_2}{\sqrt 3}\right)   \left(\begin{array}{ccc}
1&0&0\\
0&-2&0\\
0&0&1\\
\end{array}\right)
\end{split}
\end{equation}
Clearly, the embedding (\ref{emb-Bais}) is very convenient to obtain spherically symmetric 
monopoles \cite{Weinberg82}. It is also helpful to examine the fields and low-energy dynamics of 
the charge two BPS monopoles \cite{Irwin97}. Depending on the bondary conditions and 
pattern of the symmetry, some other ans\"atze can be implemented to investigate static monopole 
solutions, as, for example, the harmonic map ansatz \cite{Sutcliffe99} which was used to construct
non-Bogomol'nyi $SU(N)$ BPS monopoles.

In our consideration we shall consider  
ans\"atze for the Higgs field of a spherically symmetric $\beta_i$ monopole configuration. Depending
on the way of the $SU(2)$-embedding, it can be taken\footnote{The first of these ans\"atze 
(in a different  basis of the simple roots) was already used in \cite{Kunz87,Brihaye01}.} 
as a generalization of the the embedding (\ref{emb-Bais}) 

\begin{equation}        \label{h-1}
\begin{split}
\vec \beta_i:~~~&\Phi(r) = \Phi_1(r) \tau^{(i)}_r + \frac{\sqrt 3}{2} \Phi_2 (r) D^{(i)};\\
A_r =& 0;\quad A_\theta = [1-K(r)] \tau^{(i)}_\phi;\quad
A_\phi = -\sin\theta [1-K(r)] \tau^{(i)}_\theta\\
\end{split}
\end{equation}
where $i=1,2,3$ and we make use of the $su(2)$ matrices 
$\tau^{(i)}_r = \left(T_{(i)}^a {\hat r}^a\right)$, 
$\tau^{(i)}_\theta = \left(T_{(i)}^a {\hat \theta}^a\right)$ and 
$\tau^{(i)}_\phi = \left(T_{(i)}^a {\hat \phi}^a\right)$. The 
diagonal matrices $D^{(i)}$, which define the embedding along 
corresponding simple root, are simple  $SU(3)$ hypercharge  
$$
D^{(1)} \equiv Y = \frac{2}{\sqrt 3}H_2 = \frac{1}{3}\diag(1,1,-2),
$$ 
the $SU(3)$ electric charge operator 
$$D^{(2)} \equiv Q = T^3 + \frac{Y}{2} = H_1 + \frac{H_2}{\sqrt 3} = \frac{1}{3}\diag(2,-1,-1)$$ 
and its conjugated 
$$D^{(3)} \equiv {\tilde Q} = T^3 - \frac{Y}{2} = \frac{1}{3}\diag(1,-2,1).$$
The normalization of the  ans\"atze (\ref{h-1}) corresponds 
to the $su(3)$-norm of the Higgs field $|\Phi|^2 =  \Phi_1^2 + \Phi_2^2$ for any embedding.

Inserting the ans\"atze (\ref{h-1}) into the Lagrangian density (\ref{LagrGG-8}) yields 
\bea           \label{Lagr-comp}
-L &=& \Tr \left\{ \frac{1}{r^2}F_{r \theta}^2  + \frac{1}{r^2 \sin\theta^2} F_{r \phi}^2 
+ \frac{1}{r^4\sin^2\theta}F_{\theta\phi}^2\right\}\\
&+& \Tr \left\{ (D_r\Phi)^2 +  \frac{1}{r^2}(D_\theta\Phi)^2 +  \frac{1}{r^2 \sin\theta^2}(D_\phi\Phi)^2\right\} 
+ \lambda (|\Phi|^2 - 1)^2\nonumber\\
&=&\frac{1}{2r^4}\biggl\{ 2(r \partial_r K)^2  
+(1 - K^2)^2 \biggr\} + \frac{1}{2r^2}\biggl\{ 
(r\partial_r \Phi_1)^2 + (r\partial_r \Phi_2)^2 + 2K^2 \Phi_1^2\biggr\}\nonumber\\ 
&+&  \lambda (\Phi_1^2 + \Phi_2^2 - 1)^2,\nonumber
\eea
Straightforward variation of the Lagrangian 
${\cal L} = \int d^3 x L$ with respect to the gauge field profile functions $K(r)$ 
and the scalar field functions 
$\Phi_1(r),\Phi_2(r)$ gives the system of the coupled non-linear differential equations of second order:
   \bea           \label{Lagr-equations} 
0 &=&  \partial_r^2 K -\frac{K(K^2-1)}{r^2}  - \Phi_1^2 K =0;\\[3pt]
0 &=& 2\Phi_1 K^2 + 4\lambda r^2 \Phi_1(\Phi_1^2 + \Phi_2^2 -1) - r^2 \partial^2_r \Phi_1 
- 2r \partial_r \Phi_1;\nonumber\\
0&=& 4\lambda r^2 \Phi_2(\Phi_1^2 + \Phi_2^2 -1) - r^2 \partial^2_r \Phi_2 - - 2r \partial_r \Phi_2\nonumber
\eea
Clearly, these equations are identical for any $SU(2)$ embedding.
However the boundary conditions we have to impose on the Higgs field, 
depend from the type of the embedding.  

Let us consider the behavior of the scalar field of the configurations (\ref{h-1}) 
along positive direction of the $z$-axis. We obtain 
\begin{equation}        \label{h-1-as}
\begin{split}
\vec \beta_1:~~~&\Phi(r,\theta){\biggl| \biggr.}_{\theta=0} = \Phi_1 H_1 + \Phi_2 H_2 = (\vec h \cdot \vec H);\nonumber\\
\vec \beta_2:~~~&\Phi(r,\theta){\biggl| \biggr.}_{\theta=0} = \frac{1}{2}\left[(\sqrt 3 \Phi_2 - \Phi_1)H_1 + 
(\sqrt 3 \Phi_1 + \Phi_2) H_2\right] = (\vec h \cdot \vec H);\nonumber\\
\vec \beta_3:~~~&\Phi(r,\theta){\biggl| \biggr.}_{\theta=0} = \frac{1}{2}\left[(\sqrt 3 \Phi_2 + \Phi_1)H_1 + 
(\sqrt 3 \Phi_1 - \Phi_2) H_2\right] = (\vec h \cdot \vec H).
\end{split}
\end{equation}
That yields the components of the vector $\vec h$ which determines the nature of the symmetry breaking. 

The boundary conditions we can impose on configurations, which minimise the action (\ref{LagrGG-8})
are of different types. First, the Higgs potential vanises on the spacial asymptotic, that is as $r \to \infty$
$$
|\Phi|^2 = \Phi_1^2 + \Phi_2^2 = 1
$$ 
Second, the inner product of the   vector $\vec h$ with all roots have to be non-negative 
for any embedding. That yields 
\begin{equation}        \label{h-ort}
\begin{split}
\vec \beta_1:~~~(\vec \beta_1 \cdot \vec h) &= \Phi_1 \ge 0;\quad  (\vec \beta_2 \cdot \vec h) = -\frac{\Phi_1}{2} 
+ \frac{\sqrt 3}{2}\Phi_2 \ge 0;  \quad (\vec \beta_3 \cdot \vec h) = \frac{\Phi_1}{2} 
+ \frac{\sqrt 3}{2}\Phi_2 \ge 0\\[3pt]
\vec \beta_2:~~~(\vec \beta_1 \cdot \vec h) &= -\frac{\Phi_1}{2} 
+ \frac{\sqrt 3}{2}\Phi_2 \ge 0;  \quad (\vec \beta_2 \cdot \vec h) = \Phi_1 \ge 0; \quad (\vec \beta_3 \cdot \vec h) = 
\frac{\Phi_1}{2} + \frac{\sqrt 3}{2}\Phi_2 \ge 0\\[3pt]
\vec \beta_3:~~~(\vec \beta_1 \cdot \vec h) & = \frac{\Phi_1}{2} 
+ \frac{\sqrt 3}{2}\Phi_2 \ge 0;  \quad (\vec \beta_2 \cdot \vec h) = \frac{\Phi_1}{2} - \frac{\sqrt 3}{2}\Phi_2 \ge 0; 
\quad (\vec \beta_3 \cdot \vec h) = \Phi_1 \ge 0.
\end{split}
\end{equation}
Thirdly, the covariant derivatives of the Higgs field have to vanish at spacial infinity, that is 
\begin{equation}        \label{D-vanish}
\begin{split}
D_r \Phi &=r\partial_r \Phi_1 \tau^{(i)}_r  + \frac{\sqrt 3}{2} \partial_r \Phi_2 D^{(i)} = 0;\\
D_\theta \Phi &=(K-1)\Phi_1 \tau^{(i)}_\theta = 0;\\ 
D_\phi \Phi &=\sin\theta (K-1)K \Phi_1 \tau^{(i)}_\phi = 0   
\end{split}
\end{equation}
And finally, the solution has to be regular at the origin. The condition on the short distance
behavior implies   
$$
K(r) \to 1; \quad \Phi_1(r) \to 0; \quad \partial_r \Phi_2(r) \to 0, 
$$
as $r \to 0$. The energy density also goes to 0 in that limit. 

\section{Composite monopole solution and various limits of the symmetry breaking}

We are interested in the investigation of the properties of the configuration, which 
correspond to the embedding along the composite root $\vec \beta_3$. The physical meaning of the 
third of the ans\"atze for the scalar field (\ref{h-1}) becomes more clear if we note that 
on the spacial asymptotic this configuration really corresponds to the   
Higgs field of two distinct fundamental monopoles, 
$(1,0)$ and $(0,1)$. Indeed, outside of the cores of these monopoles in the Abelian gauge the 
scalar field can be written as superposition:
$$
\Phi (r\to \infty) =  v_1 T_{(1)}^3 + v_2 T_{(2)}^3 =  \frac{1}{2}\left(\begin{array}{ccc}
v_1&0&0\\
0&v_2-v_1&0\\
0&0&-v_2\\
\end{array}\right)
$$
where the Higgs field of the $\beta_1$ and $\beta_2$ monopoles is taking the  
vacuum values $ v_1, v_2$ respectively.  
 
Rotation of this configuration by the matrices of the $SU(2)$ subgroup 
which is defined by the third composite root $\vec \beta_3$ 
$$
U =  \left(\begin{array}{ccc}
\cos \frac{\theta}{2}&0&\sin \frac{\theta}{2} e^{-i\phi}\\
0&1&0\\
-\sin \frac{\theta}{2} e^{i\phi}&0&\cos \frac{\theta}{2}\\
\end{array}\right)
$$  
yields 
\begin{equation}  \label{two-higgs-rot}
U^{-1} \Phi U =  \frac{1}{2}[v_1 + v_2] \tau^{(3)}_r +  
\frac{3}{4} [v_1 - v_2] {\tilde Q}
\end{equation}
Up to the obvious reparametrization of the  
shape functions of the scalar field 
\begin{equation} \label{two-higgs}
\Phi_1 \to \frac{1}{2} \left(F_1(r)+F_2(r)\right);\qquad 
\Phi_2 \to \frac{\sqrt 3}{2} \left(F_1(r)-F_2(r)\right) 
\end{equation}
where the functions $F_1,F_2$ have the vacuum expectation values $v_1,v_2$ respectively, 
the configuration (\ref{two-higgs}) precisely corresponds to the third of the ans\"atze 
(\ref{h-1}). Note that because the $su(3)$-norm of the scalar field is set to be unity, 
the vacuum values must satisfy the condition  
$$ 
v_1^2 + v_2^2 - v_1v_2 =1
$$
Moreover, the reparametrization  (\ref{two-higgs}) allows to write 
the scalar field of the $\beta_3$ monopole  
along positive direction of the $z$-axis as  
\begin{equation} \label{beta3-rep}       
\vec \beta_3:~~~\Phi(r\to \infty,\theta){\biggl| \biggr.}_{\theta=0} = \left(v_1 - \frac{v_2}{2}
\right)H_1 +  
\frac{\sqrt 3}{2}v_2 H_2 = (v_1 \vec \beta_1 + v_2 \vec \beta_2)\cdot \vec H = (\vec h \cdot \vec H)
\end{equation}
Thus,  the asymptotic values  $v_1$ and $v_2$ 
are the coefficients of the expansion of the vector $\vec h$ in the basis of the simple 
roots and on the spacial asymptotic the fields $F_1 (\vec \beta_1 \cdot \vec H)$ and 
$ F_2 (\vec \beta_2\cdot \vec H)$ can be identified with the  
Higgs fields of the first and second fundamental monopole respectively. 

The solution of the equations (\ref{Lagr-equations}) becomes very simple in the BPS limit.  
Then the third equation is decoupled and its solution, which is regular at the origin, 
is just a constant $\Phi_2 = C$, $C \in [0;1]$. The shape functions of the scalar and 
gauge field are well known rescaled Bogomol'nyi solutions 
\begin{equation}   \label{solution-BPS}
K(r') = \frac{r'}{\sinh r'};\qquad \Phi_1(r') = \sqrt{1-C^2}\, \coth r' - \frac{1}{r'},
\qquad {\rm where}~~~ r' =  r \sqrt{1-C^2}  
\end{equation}
with a long-range field $\Phi_1$. 

Now we may treat the configuration, which corresponds to the 
minimal $SU(3)$ symmetry breaking, as a special case of maximal symmetry breaking, namely we can  
start from an abritrary orientation of the vector $\vec h$ which is compatible with the 
boundary conditions (\ref{h-ort}). 

Embedding along the composite simple root $\vec \beta_3$ gives two fundamental monopoles, which 
in the case of the maximal symmetry breaking, are charged with respect to different $U(1)$ 
subgroups and are on top of each other. The configuration with a minimal energy corresponds 
to the  boundary condition  $(\Phi_1)_{vac} =1, (\Phi_2)_{vac}=0$. 
We can interpret it, by making use of 
Eq. (\ref{two-higgs}), as two identical monopoles of the same mass. This degeneration is lifted as 
the value of the constant solution $\Phi_2=C$ increases, the vector of the Higgs field  $\vec h$ 
smoothly rotates in the root space and the boundary 
conditions begin to vary. 

According to the parametrization (\ref{two-higgs}) increasing of the 
constant $C$ results in the splitting of the vacuum values of the 
scalar fields of the first and second fundamental monopoles, the 
$\beta_1$-monopole is getting heavier than the $\beta_2$-monopole. Note that the 
shape function of the vector field remains almost unaffected by this variation of the boundary
conditions as shown in Fig \ref{f-09}.  

One would expect that in the limiting case of the minimal symmetry breaking 
the $\beta_2$ monopole is getting massless, that is in that limit the vacuum value 
of the field $F_2$ should vanish, $v_2 \to 0$. However 
two monopoles are overlapped and the presence of the massive monopole changes the situation.  
Indeed, the symmetry outside the core of the $\beta_1$-monopole is breaked down to $U(1)$ 
which also changes the pattern of the symmetry breaking by the scalar field of the second monopole. 
One can see that the  vector  $\vec h$  becomes orthogonal to the simple 
root $\vec \beta_2$ when $(\Phi_1)_{vac} = \frac{\sqrt 3}{2}, (\Phi_2)_{vac} = \frac{1}{2}$, or 
$v_2 = \frac{v_1}{2}$. Going back to the Eg. (\ref{beta3-rep}) we can see that in this case 
on the spacial asymptotic the scalar field  along $z$ axis is  
$$
\Phi(r\to \infty,\theta){\biggl| \biggr.}_{\theta=0} = \frac{3}{4} v_1 D^{(2)}  
$$
where $D^{(2)} = H_1 + \frac{1}{\sqrt 3} H_2$. Thus, the symmetry 
is still maximally brocken and both monopoles are massive. 

Equation (\ref{beta3-rep}) indicates that the symmetry is minimally brocken if 
the vector  $\vec h$  becomes orthogonal to the simple 
root $\vec \beta_1$  and $v_1 = \frac{v_2}{2}$. 
Then the eigenvalues of the scalar field are 
the same as $H_2$, that is the unbrocken symmetry group is really 
$U(2)$. 
However for the third composite root such a situation corresponds to the negative value of the 
inner product $(\vec h\cdot \vec \beta_2)$ and it has to be excluded. Thus, the maximal 
vacuum value of the second component of the Higgs field of $\beta_3$-monopole is 
$(\Phi_2)_{vac} = \frac{1}{2}$. This is a border value which, according to (\ref{h-ort}), 
separates the composite $\beta_3$-monopole from a single fundamental $\beta_i$-monopole, 
for which $(\vec h\cdot \vec \beta_i) \ge 0$ if $(\Phi_2)_{vac} \ge \frac{1}{2}$. In that 
case the pattern of symmetry breaking becomes more simple.  
Futher increasement of the vacuum value $(\Phi_2)_{vac}$ smoothly moves the configuration to the  
limit $\Phi_1 \to 0$. Then the vector $\vec h$ becomes orthogonal to one of the simple roots that 
is the gauge symmetry is brocken minimally and this is the case of a 
''massless'' monopole. 

In a general case of non-zero scalar coupling $\lambda$ the system of equations 
(\ref{Lagr-equations}) may be solved numerically. This calculation have been done in 
\cite{Kunz87} for the all range of possible boundary conditions. 
Here we only make another interpretation of the results, which is related with 
separation of the single fundamental monopole from a composite one. 
 
As in the paper \cite{Kunz87}, we performed the calculation using COLSYS 
package. The profile function of the gauge and scalar field are plotted for several 
values of the  vacuum expectation value $(\Phi_2)_{vac}=C$ in Figures \ref{f-02}-\ref{f-09}.
In the following table, we summarize our results of the evaluation of the mass of the 
configuration in units of $4\pi$ as a function of the boundary conditions 
on the  vacuum value $(\Phi_2)_{vac}=C$ for values $\lambda = 0$ (BPS monopole), and $\lambda= 1, 10$:
\begin{center}
\begin{tabular}{|l|c|c|c|c|}\hline  
Embedding & $\quad$ $C$ $\quad$& $\quad$ $M{\biggl| \biggr.}_{\lambda=0}$ $\quad$  
          &  $\quad$ $M{\biggl| \biggr.}_{\lambda=1}$ $\quad$  
          & $\quad$ $M{\biggl| \biggr.}_{\lambda=10}$ $\quad$\\[4pt] \hline
Single $\beta_i$-monopole     &1.0    & 0.0    & 0.0  & 0.0 \\ \hline
                              &0.9999 & 0.011 & 0.011 & 0.011 \\ \hline
                              &0.999  & 0.042 & 0.042 & 0.042 \\ \hline
                              &0.99   & 0.138 & 0.139 & 0.140 \\ \hline
                              &0.95   & 0.309 & 0.314 & 0.317 \\ \hline
                              &0.90   & 0.433 & 0.456 & 0.472 \\ \hline
                              &0.80   & 0.597 & 0.634 & 0.682 \\ \hline
                              &0.70   & 0.711 & 0.777 & 0.820\\ \hline
                              &0.60   & 0.797 & 0.897 & 0.971 \\ \hline 
                              &0.50   & 0.863 & 1.001 & 1.034\\ \hline
                                                  \hline
Composite $\beta_3$-monopole  &0.50  &  0.863 & 1.001 & 1.034\\ \hline
                              &0.40  &  0.913 & 1.090 & 1.121 \\   \hline
                              &0.30  &  0.951 & 1.166 & 1.217\\ \hline
                              &0.20  &  0.976 & 1.229 & 1.306\\   \hline
                              &0.10  &  0.992 & 1.273 & 1.390 \\  \hline
                              &0.05  &  0.996 & 1.284 & 1.423\\  \hline
                              &0.    &  1.00 & 1.291 & 1.467\\ \hline
\end{tabular}
\end{center}
The calculations shows that 
the energy of interaction of two distinct fundamental non-BPS monopoles 
of the same mass ($(\Phi_2)_{vac}= 0$), which are on top of each other, is very strong 
and this configuration seems to be an absolute minimum of the energy functional. 

One may see that in the another limiting case $(\Phi_2)_{vac}= 1/2$ 
the mass degeneration between a single fundamental monopole and composite monopole 
is preserved at finite scalar coupling. Note that a single fundamental monopole 
becomes massless if the profile function of the vector  field is constant 
everywhere: $K= 1$. This is a case of the minimal symmetry breaking.  

\subsection*{Minimal symmetry breaking solution} 
Let us consider behavior of a single fundamental monopole solution as 
the  vacuum expectation value $(\Phi_2)_{vac}$ approaches the limit $C=1$. 
Fugure \ref{f-06} shows that the monopole is spreading out in space as it was expected.    
First, we note that in this limit the potential of the scalar field $V(\Phi)$ 
vanishes everywhere for any value of $\lambda$, thus it is a BPS-like configuration.
The `hedgehog' component $\Phi_1$ vanishes while the second component remains a constant: 
$\Phi_2 = H_2$, thus this configuration becomes topologically trivial and can be continuously 
deformed into thrivial solution $K= 1$.       

Indeed, the energy, which corresponds to (\ref{Lagr-comp}), in the limit of the
minimal symmetry breaking becomes simple 
\begin{equation} \label{pseudo-kink}
E = 4\pi \int dr \left\{ (\partial_r K)^2 + \frac{(1-K^2)^2}{2 r^2}\right\}    
\end{equation}
Moreover, the covariant derivatives of the Higgs field (\ref{D-vanish}) vanish everywhere 
while the non-Abelian magnetic field becomes 
\begin{equation}
B_r = \frac{1-K^2}{r^2}\, \tau_r^{(i)};\qquad  
B_\theta = \frac{\partial_r K}{r}\, \tau_\theta^{(i)};\qquad
B_\phi = -\frac{\partial_r K}{r}\, \tau_\phi^{(i)}
\end{equation} 
that is the radial derivative ${\partial_r} K$ is not 
vanishing as $r \to \infty$. It would define 
the magnitude of the angular components $B_\theta \sim  B_\phi \sim r^{-1}$ which would 
appear beside the radial Coulomb field $B_r \sim r^{-2}$.  However the topological charge 
$g = \Tr\, \int d^2 S B_n \Phi = 0$ since $\Phi = H_2$.  

If we identify the term $\frac{(1-K^2)^2}{2 r^2}$ as a new scalar ``potential'' of the field $K(r)$,  
the energy functional (\ref{pseudo-kink}) gets some simularity with the 
one-dimensional $\phi^4$ theory. However the kink solution of the latter model interpolate 
between two-fold degenerated vacua $\phi_{vac} = \pm 1$, 
while the potential of the former model vanishes on the spacial 
asymptotic and the profile function $K(r)$ would interpolate between $K(0) =1$ and 
$K(r\to \infty) \to 0$ and such a configuration is instable. Indeed, let us consider 
the small spherically symetric fluctuations of the fields about the configuration 
with minimal symmetry breaking:
$$
K(r) \to K(r) + a(r);\qquad \Phi_1(r) \to \chi(r);\qquad  \Phi_2(r) \to 1 + \eta(r)
$$  
Then expansion of the energy in terms of these variations yelds 
the operator of second derivatives 
\begin{equation}
{\cal D}^{(2)} = \left(\begin{array}{ccc}
-\partial_r^2 + \frac{3K^2 -1}{r^2} &0&0\\
0& -\frac{1}{2} \partial_r^2 + \frac{K^2}{r^2} &0\\
0&0&-\frac{1}{2} \partial_r^2 + 4 \lambda \\
\end{array}\right)
\end{equation}
which is diagonal and has eigenfunctions 
$$
 {\cal D}^{(2)} 
\begin{pmatrix}a\\
r \chi\\
r \eta
\end{pmatrix} = - \omega^2 \begin{pmatrix}a\\
r \chi\\
r \eta
\end{pmatrix}
$$
Each negative eigenvalue $-\omega^2$ generates an exponentially 
growing instability of the configuration. One may easily see that this equation 
has a number of negative modes in the region $r \to \infty$. 

Thus, in the limiting case of the minimal symmetry breaking a single isolated 
fundamental monopole is dissolving into the topologically trivial sector  
and only a constant component of the scalar field $\Phi_2 = H_2$ survives.      

%Concluding, we have discussed the properties of non-BPS $SU(3)$ monopoles 
%depending on the boundary conditions on the Higgs field. 
\bigskip

{\bf Acknowledgements}
\medskip

I am very grateful to Jutta Kunz for invaluable help and support. All the numerical 
calculations were performed with her help. I would like to acknowledge the hospitality 
at the Abdus Salam
International Center for Theoretical Physics where this work was completed (ICTP
Preprint, IC/2003/180).

\newpage
%%%%%%%%%%%%%%%%%%%%%%%%%%%%%%%%%%%%%%%%%%%%%%%%%%%%%%%%%%%%%%%%%%%%%%
%%%%%%%%%%%%%%%%%%%%%%%%%%%%%%%%%%%%%%%%%%%%%%%%%%%%%%%%%%%%%%%%%%%%%
\begin{figure}[t]
\begin{center}
\setlength{\unitlength}{1cm}
\lbfig{f-03}
\begin{picture}(13,7.5)
\put(2.0,-0.5)
{\mbox{
\psfig{figure=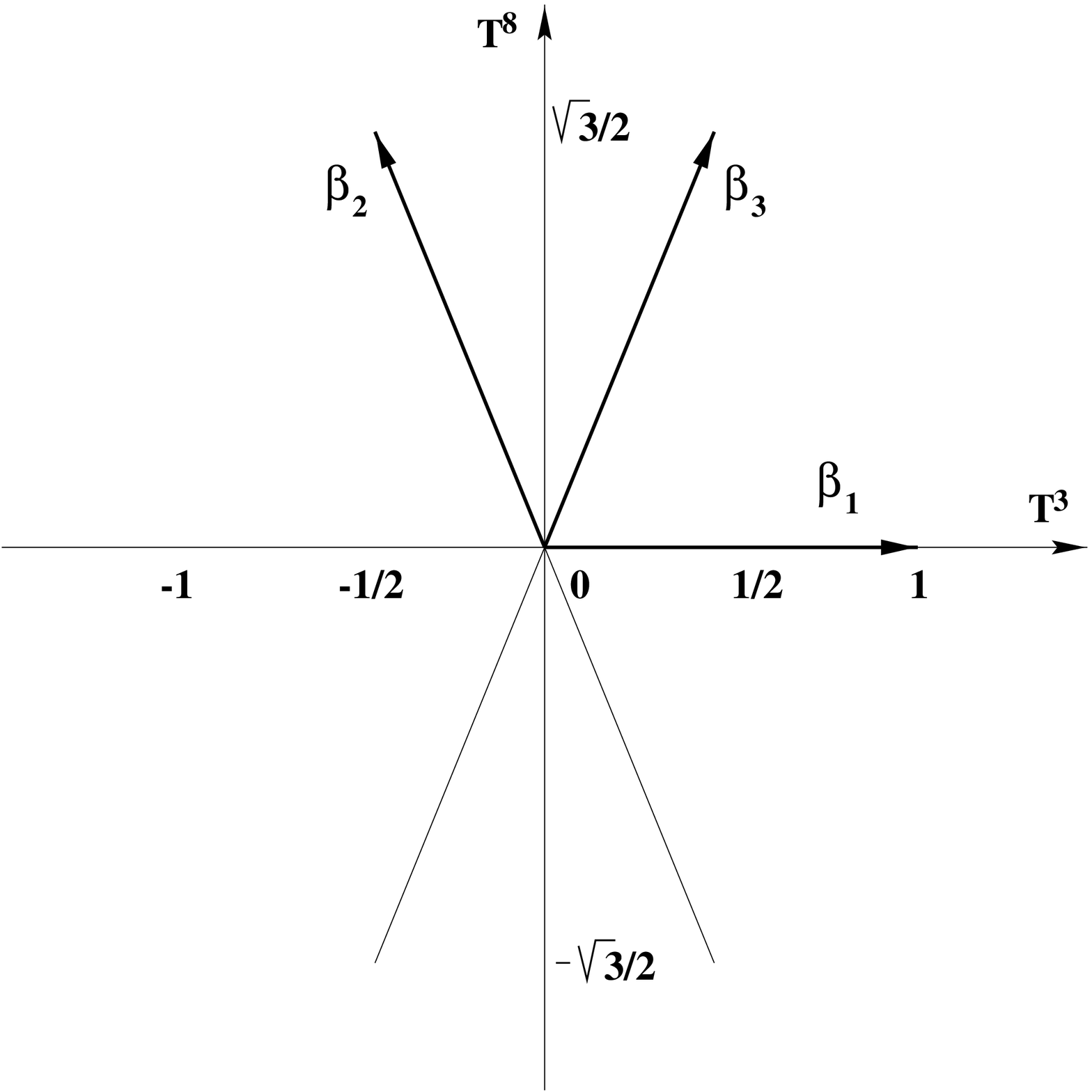,height=8.0cm}}}
\end{picture}
\caption{
$SU(3)$ simple root basis.}
\end{center}
\end{figure}
%%%%%%%%%%%%%%%%%%%%%%%%%%%%%%%%%%%%%%%%%%%%%%%%%%%%%%%%%%%%%%%%%%%%%%%%
%%%%%%%%%%%%%%%%%%%%%%%%%%%%%%%%%%%%%%%%%%%%%%%%%%%%%%%%%%%%%%%%%%%%%% 
\begin{figure}[thb]
\begin{center}
\setlength{\unitlength}{1cm}
\lbfig{f-02}
\begin{picture}(13,8.0)
\put(0.0,7.8)
{\mbox{
\psfig{figure=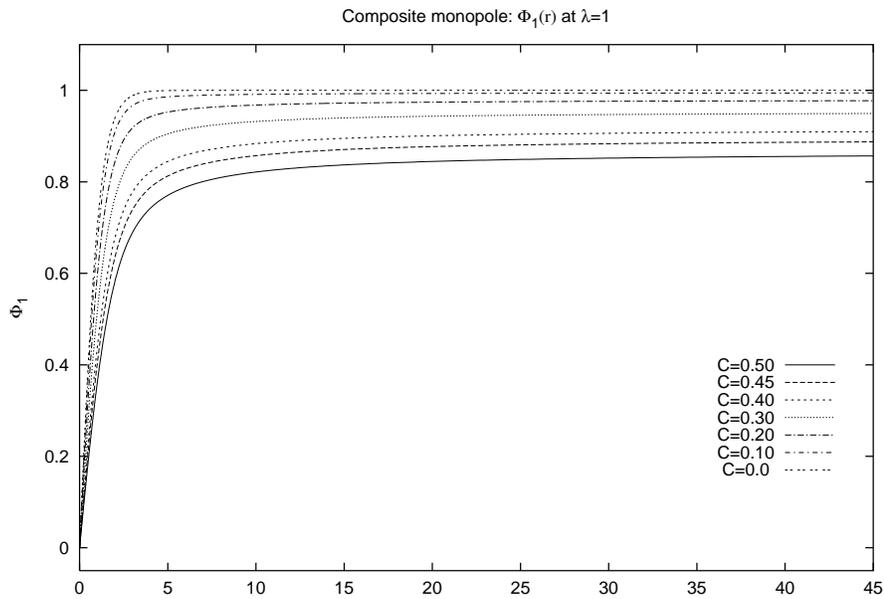,height=12.0cm, angle =-90}}} 
\end{picture} 
\caption{
Structure functions of the Higgs field component $\Phi_1(r)$ of the 
single fundamental monopole for different boundary conditions on the 
second component $\Phi_2(r)$  ($\lambda =1$).    
}
\end{center} 
\end{figure} 
%%%%%%%%%%%%%%%%%%%%%%%%%%%%%%%%%%%%%%%%%%%%%%%%%%%%%%%%%%%%%%%%%%%%%
%%%%%%%%%%%%%%%%%%%%%%%%%%%%%%%%%%%%%%%%%%%%%%%%%%%%%%%%%%%%%%%%%%%%%% 
\begin{figure}[thb]
\begin{center}
\setlength{\unitlength}{1cm}
\lbfig{f-05}
\begin{picture}(13,8.0)
\put(0.0,7.8)
{\mbox{
\psfig{figure=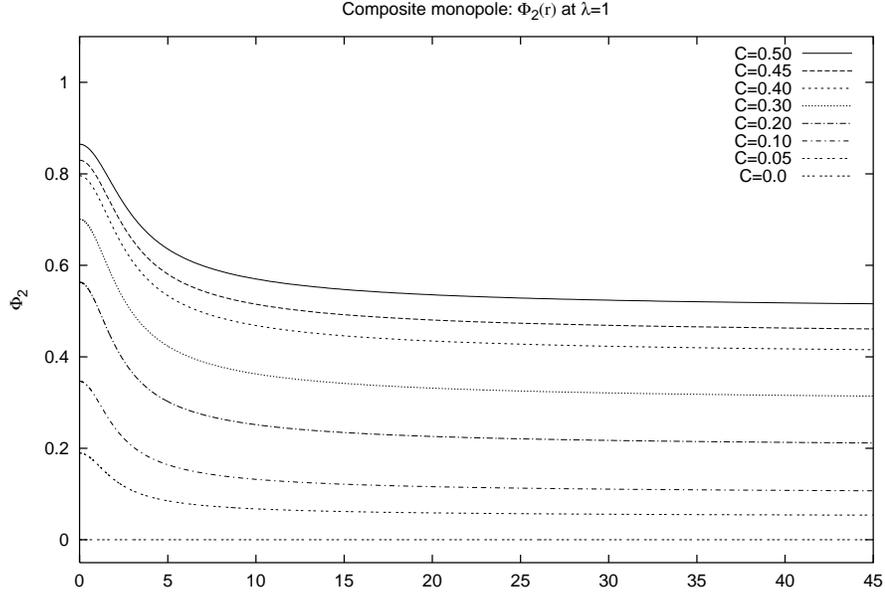,height=12.0cm, angle =-90}}} 
\end{picture} 
\caption{
Structure function of the Higgs field component $\Phi_2(r)$ of the 
single fundamental monopole with different boundary conditions ($\lambda =1$).    
}
\end{center} 
\end{figure} 
%%%%%%%%%%%%%%%%%%%%%%%%%%%%%%%%%%%%%%%%%%%%%%%%%%%%%%%%%%%%%%%%%%%%%
%%%%%%%%%%%%%%%%%%%%%%%%%%%%%%%%%%%%%%%%%%%%%%%%%%%%%%%%%%%%%%%%%%%%%% 
\begin{figure}[thb]
\begin{center}
\setlength{\unitlength}{1cm}
\lbfig{f-06}
\begin{picture}(13,8.0)
\put(0.0,7.8)
{\mbox{
\psfig{figure=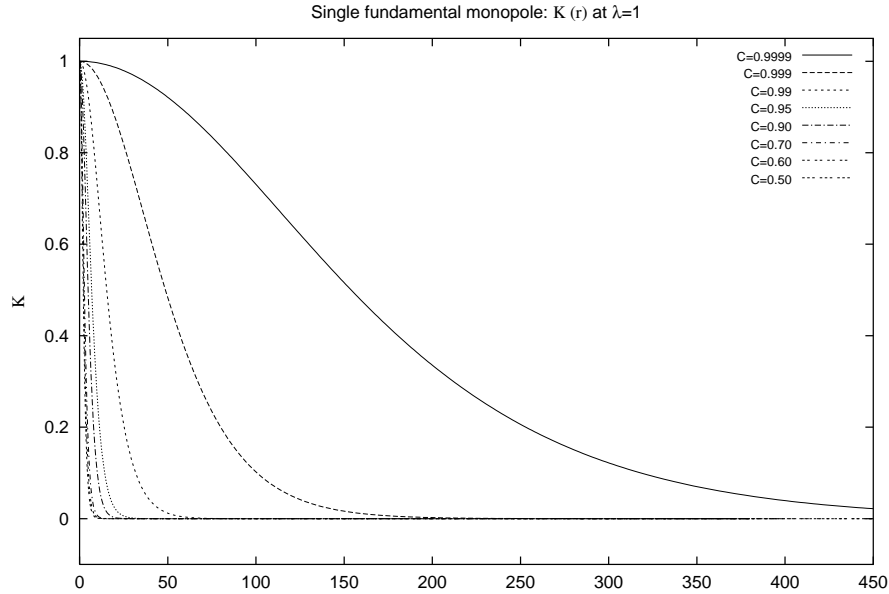,height=12.0cm, angle =-90}}} 
\end{picture} 
\caption{
Structure function of the gauge field $K(r)$ of the 
single fundamental monopole for different boundary conditions on the
second component of the Higgs field $\Phi_2$. The monopole 
is spreading out as $\Phi_2(r\to \infty)$ is approaching the limit $C=1$ which 
corresponds to the minimal symmetry breaking. 
}
\end{center} 
\end{figure} 
%%%%%%%%%%%%%%%%%%%%%%%%%%%%%%%%%%%%%%%%%%%%%%%%%%%%%%%%%%%%%%%%%%%%%
%%%%%%%%%%%%%%%%%%%%%%%%%%%%%%%%%%%%%%%%%%%%%%%%%%%%%%%%%%%%%%%%%%%%%% 
\begin{figure}[thb]
\begin{center}
\setlength{\unitlength}{1cm}
\lbfig{f-07}
\begin{picture}(13,8.0)
\put(0.0,7.8)
{\mbox{
\psfig{figure=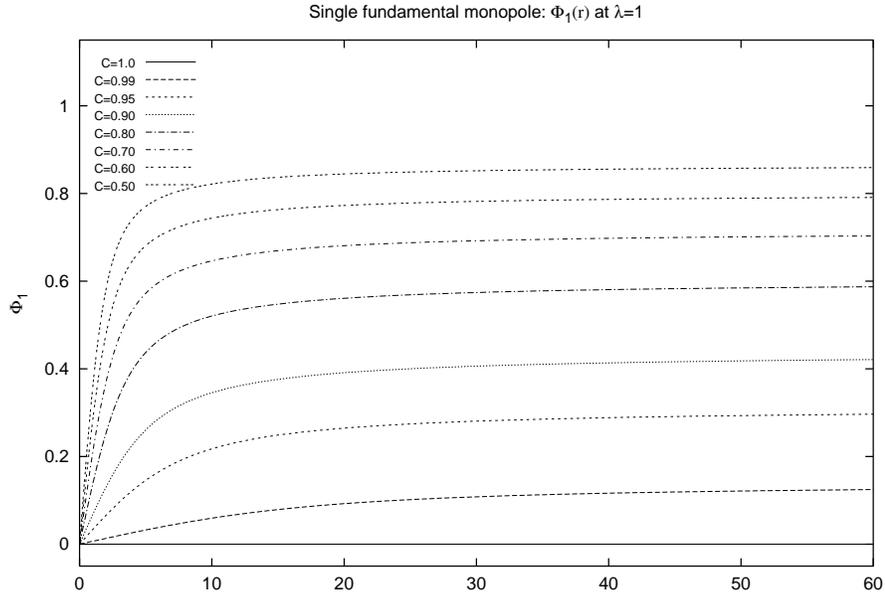,height=12.0cm, angle =-90}}} 
\end{picture} 
\caption{
Structure function of the Higgs field component $\Phi_1(r)$ of the 
composite monopole for different boundary conditions on the 
second component $\Phi_2(r)$ ($\lambda =1$).    
}
\end{center} 
\end{figure} 
%%%%%%%%%%%%%%%%%%%%%%%%%%%%%%%%%%%%%%%%%%%%%%%%%%%%%%%%%%%%%%%%%%%%%
%%%%%%%%%%%%%%%%%%%%%%%%%%%%%%%%%%%%%%%%%%%%%%%%%%%%%%%%%%%%%%%%%%%%%% 
\begin{figure}[thb]
\begin{center}
\setlength{\unitlength}{1cm}
\lbfig{f-08}
\begin{picture}(13,8.0)
\put(0.0,7.8)
{\mbox{
\psfig{figure=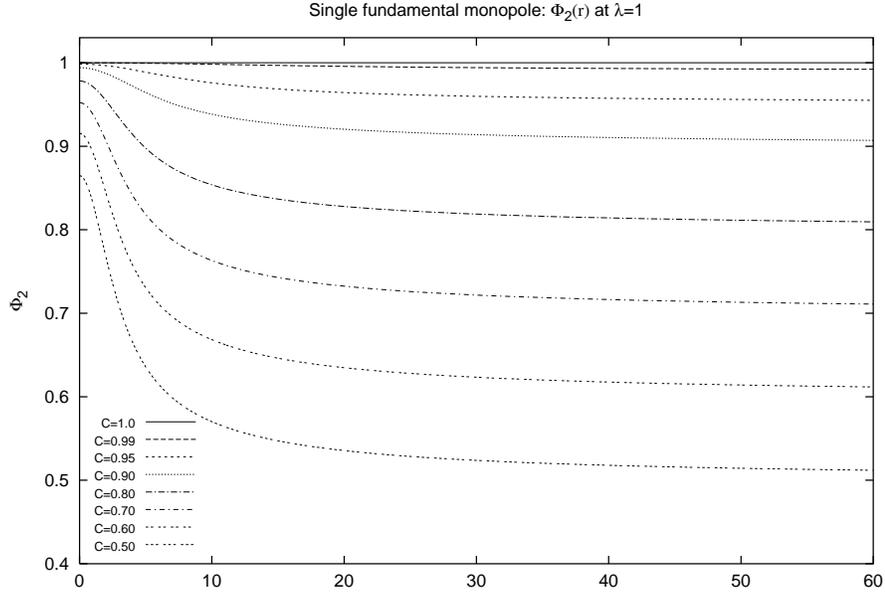,height=12.0cm, angle =-90}}} 
\end{picture} 
\caption{
Structure functions of the Higgs field component $\Phi_2(r)$ of the 
composite monopole for different boundary conditions ($\lambda =1$).    
}
\end{center} 
\end{figure} 
%%%%%%%%%%%%%%%%%%%%%%%%%%%%%%%%%%%%%%%%%%%%%%%%%%%%%%%%%%%%%%%%%%%%%
%%%%%%%%%%%%%%%%%%%%%%%%%%%%%%%%%%%%%%%%%%%%%%%%%%%%%%%%%%%%%%%%%%%%%% 
\begin{figure}[thb]
\begin{center}
\setlength{\unitlength}{1cm}
\lbfig{f-09}
\begin{picture}(13,8.0)
\put(0.0,7.8)
{\mbox{
\psfig{figure=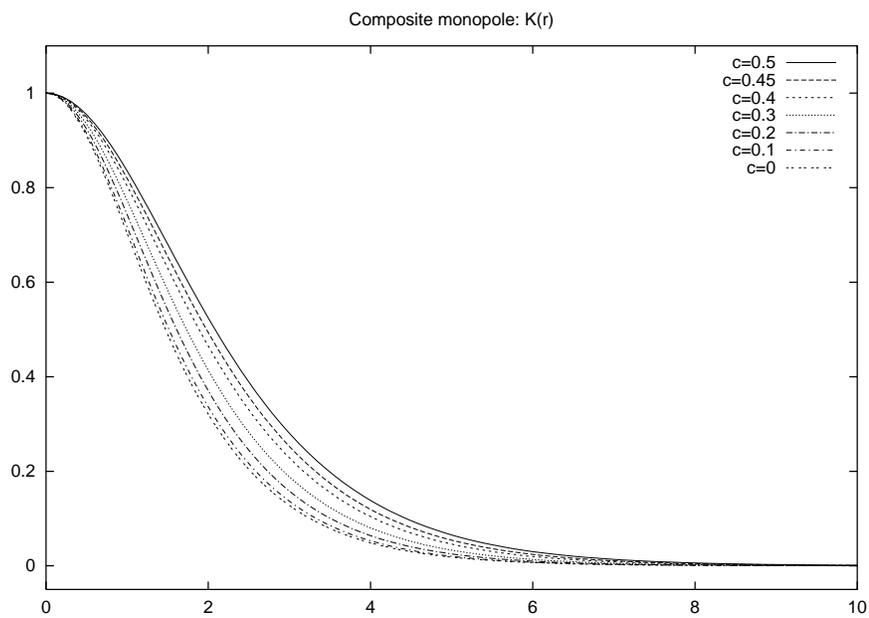,height=12.0cm, angle =-90}}} 
\end{picture} 
\caption{
Structure functions of the gauge field $K(r)$ of the 
composite monopole for different boundary conditions on the
second component of the Higgs field $\Phi_2$ ($\lambda =1$).    
}
\end{center} 
\end{figure} 
%%%%%%%%%%%%%%%%%%%%%%%%%%%%%%%%%%%%%%%%%%%%%%%%%%%%%%%%%%%%%%%%%%%%%

\end{document}